\documentclass[draft,jgrga]{agutex}

  \usepackage[dvipdf]{graphicx}

  \setkeys{Gin}{draft=false}
\authorrunninghead{YERMOLAEV ET AL.}

\titlerunninghead{DYNAMICS SOLAR-WIND STREAMS}

\begin{document}

%
%

\title{Dynamics of large-scale solar-wind streams obtained by the double superposed epoch analysis}
%
%

%
%




\authors{Yu. I. Yermolaev, \altaffilmark{1}
I. G. Lodkina, \altaffilmark{1} 
N. S. Nikolaeva , \altaffilmark{1} 
M. Yu. Yermolaev \altaffilmark{1}
}

\altaffiltext{1}{Space Plasma Physics Department, Space Research Institute, 
Russian Academy of Sciences, Profsoyuznaya 84/32, Moscow 117997, Russia. 
(yermol@iki.rssi.ru)}






%
%


\begin{abstract} 

Using the OMNI data for period 1976-2000 we investigate the temporal profiles of 
20 plasma and field parameters in the disturbed large-scale types of solar wind (SW): 
CIR, ICME (both MC and Ejecta) and Sheath as well as the interplanetary shock (IS). 
To take into account the different durations of SW types we use 
the double superposed epoch analysis (DSEA) method: re-scaling the duration of the interval 
for all types in such a manner that, respectively, beginning and end for all intervals of selected type 
coincide. 
As the analyzed SW types can interact with each other and change parameters as a result of such interaction, 
we investigate separately 8 sequences of SW types:
(1) CIR, 
(2) IS/CIR,
(3) Ejecta,
(4) Sheath/Ejecta,
(5) IS/Sheath/Ejecta, 
(6) MC,
(7) Sheath/MC, and 
(8) IS/Sheath/MC. 
The main conclusion is that the behavior of parameters in Sheath and in CIR are very similar 
both qualitatively and quantitatively.
Both the high-speed stream (HSS) and the fast ICME play a role of pistons 
which push the plasma located ahead them. 
The increase of speed in HSS and ICME leads at first to formation of compression regions
(CIR and Sheath, respectively), and then to IS. The occurrence of compression regions and IS 
increases the probability of growth of magnetospheric activity. 

\end{abstract}

%
%

%

\begin{article}

%
%

\section{Introduction} 

Studying the large--scale structure of the solar wind (SW) plays the key role in the space weather investigations. 
As well known the main source of magnetospheric disturbances including the magnetic storms is 
the southward component $B_z$ of interplanetary magnetic field (IMF) 
\citep{Dungey1961,FairfieldCahill1966,RostokerFalthammar1967,Russelletal1974,Burtonetal1975,Akasofu1981}.
In the steady types of the SW streams the IMF lies near the ecliptic plane and these types are not geoeffective. 
Only disturbed types of SW streams can contain the IMF component perpendicular to the ecliptic plane and 
in particular the southward IMF component. 
Such disturbed types are the following SW streams: 
interplanetary manifestation of coronal mass ejection (ICME) including magnetic cloud (MC) and Ejecta, 
Sheath - compression region before ICME, and 
corotating interaction region (CIR) - compression region before high-speed stream (HSS) of solar wind
(see reviews and recent papers by  
\cite{Gonzalezetal1999,HuttunenKoskinen2004,YermolaevYermolaev2006,YermolaevYermolaev2010,Zhangetal2007,Yermolaevetal2012,Hietalaetal2014,Cidetal2014,Katusetal2015}
and references therein). 
In the mentioned above works the SW types are considered as sources of magnetospheric disturbance, 
i.e. the selection of SW events is carried out in connection with magnetospheric activity. 
There is only a small number of articles in which authors investigate the disturbed types of SW  streams 
without their relation with the magnetospheric activity (see, e.g. papers by  
\cite{ZurbuchenRichardson2006,Yermolaevetal2009,BorovskyDenton2010,ThatcherMuller2011,RichardsonCane2012,MitsakouMoussas2014,WuLepping2015}
).
To understand geoeffectiveness of various types of SW streams, it is necessary to compare 
the characteristics of the streams inducing magnetic storm 
with the characteristics of all events of this type 
independently of possibility to storm generation.
In the present work we analyze full sets of various solar wind types for interval 1976-2000.

As showed by numerous researches, for the majority of tasks it is not enough to analyze separate values of parameters, 
and it is necessary to study their dynamics. 
The analysis of time evolution in the interplanetary parameters using the superposed
epoch analysis (SEA) method is more informative.
The choice of zero (reference) time for SEA is important
and substantially influences the results ( 
\cite{Yermolaevetal2007b,Ilieetal2008}
). In the most part of the previous papers, 
the authors use the beginning or end of intervals as zero time for SEA 
but this choice is convenient only for studying
the beginning or the end of the interval, respectively, 
because various SW types have different durations (see, e.g., papers by  
\cite{Yermolaevetal2007b,GuptaBadruddin2009}
and references therein). 
For the analysis of dynamics of parameters in the intervals with different durations 
the methods are used in which there are two reference times and 
the initial and  final points of intervals are combined with these times, 
and points between the ends of intervals are transformed by some procedure. 
For example, in the work of 
\cite{Simmsetal2010} 
the measured points only near two reference times 
are processed by the SEA method, and points between them are not processed and ignored.
In other works (see, e.g., papers by 
\cite{YermolaevYermolaev2010,Kilpuaetal2013,Katusetal2013,Katusetal2015,Hietalaetal2014})
the durations of all intervals are made equal by artificial change of distance between points.
We use the "double" (with two reference times) SEA method (
\cite{YermolaevYermolaev2010}
),  
that is, we re-scale the duration of the interval 
of all SW types in such a manner that, respectively, beginning and end 
for all intervals of selected SW type coincide.

In the present paper we analyze the full set of disturbed SW types using one method. 
In addition to average values of parameters we analyze the dynamics of streams 
by the double superposed epoch analysis.
As CIR and Sheath are result of interaction of different types of streams, we analyze following consequences of streams: 
(1) CIR, 
(2) IS/CIR,
(3) Ejecta,
(4) Sheath/Ejecta,
(5) IS/Sheath/Ejecta, 
(6) MC,
(7) Sheath/MC,
(8) IS/Sheath/MC.
The organization of the paper is as follows: Section 2 describes data and method. 
In Section 3, we present results on dynamics of parameters in various SW types. 
Section 4 summarizes the results.

\section{Methods} 

  
We use the 1-h interplanetary plasma and magnetic field data of OMNI database 
\citep{KingPapitashvili2004} 
as a basis for our investigations.
We made our own data archive including OMNI
data and calculated (using OMNI data) additional parameters
including thermal and dynamic pressures, plasma $\beta-$ 
parameters (ratio of thermal and magnetic field pressures),
ratio of measured temperature and temperature estimated on
the basis of average velocity-temperature relation and others.
Using threshold criteria for key parameters of SW and
IMF (velocity, temperature, density, ratio of thermal to magnetic
pressure, magnitude and orientation of magnetic field,
etc.), we defined corresponding large-scale SW types and
the possible error of this identification for every 1-h point
of the archive during 1976--2000 (see paper by 
\cite{Yermolaevetal2009}, 
and site ftp://ftp.iki.rssi.ru/pub/omni/). Our
identification of SW types is based on methods similar to
ones described in many papers (see reviews by 
\cite{ZurbuchenRichardson2006,WimmerSchweingruberetal2006,Tsurutaniaetal2006}
and references therein) and basically agrees with the results of
other authors, but in contrast with other similar studies, we
used a general set of threshold criteria for all SW types and
made the identification for each 1-h point.

As we noted in Introduction, the purpose of our work is the analysis of dynamics of parameters 
in the non-stationary disturbed types of the solar wind. 
Therefore here we analyze only 4 types of the solar wind: CIR, Sheath, MC and Ejecta. 
We selected only such events for which the SW type and the edges of an interval of the type 
could be defined on the basis of measurements 
(measurements of some parameters on this interval could be absent). 
Such events was 695 for Ejecta, 451 for CIR, 402 for Sheath, and 60 for MC. 
Because of further selection on adjacent SW types the subsets of events have lower statistics, 
the smallest statistics (9 events) was for MC without Sheath and IS (see figure 6), 
and in other cases the subsets included from 18 to 372 events.
As each type has characteristic (average) duration (see, e.g., 
\cite{Jianetal2008,Yermolaevetal2009,MitsakouMoussas2014}
 and references therein), 
and duration of separate events can differ from average, 
the most effective way of studying of dynamics of parameters is 
the method of double superposed epoch analysis (DSEA) 
\citep{Yermolaevetal2010}. 
We used various fixed durations for various types of streams: 20 h for CIR, 
25 h for Ejecta and MC, 14 h for Sheath before Ejecta and 10 h for Sheath before MC. 
The time between beginning and end of interval for each event was re-scaled 
(proportionally increased/decreased) such a way that after this transformation 
all events of separated type of streams have equal durations 
in the new time reference frame.

To consider influence 
of both the surrounding undisturbed solar wind, 
and the interaction of the disturbed types of the solar wind on the parameters, 
we separately analyze the following sequences of the phenomena:
 
(1) SW/CIR/SW, 

(2) SW/IS/CIR/SW,

(3) SW/Ejecta/SW,

(4) SW/Sheath/Ejecta/SW,

(5) SW/IS/Sheath/Ejecta/SW, 

(6) SW/MC/SW,

(7) SW/Sheath/MC/SW,

(8) SW/IS/Sheath/MC/SW (see Figures 1-8). 

Parameters for undisturbed SW intervals 
(before and after disturbed types) 
are calculated using the standard (without re-scaling duration) SEA method 
on 6 points with reference points on the edges of corresponding disturbed types of stream.    
Thus, though the method used by us is similar to the former DSEA (double superposed epoch analysis) method
\citep{Yermolaevetal2010}, 
this method is more developed and can be called as Multiple superposed epoch analysis (MSEA) method.


\section{Results}

Results are presented in figures 1-8 which have similar structure 
and show the following parameters: 

a) the ratio of thermal and magnetic pressures ($\beta-$ parameter), 
the thermal pressure $P_t$, 
the ratio of measured and expected temperatures $T/T_{exp}$;

b) the proton temperature $T_p$; 

c) the solar wind velocity angles $\phi$ and $\theta$;

d) the z-component of IMF $B_z$ and y-component of interplanetary electric field $E_y$;

e) the measured and density-corrected $D_{st}$ and $D^*_{st}$ indexes; 

f) the magnitude of IMF $B$, the dynamic pressure $P_d$;

g) the y- and x-components of IMF ($B_y$ and $B_x$); 

h) the sound and alfvenic velocities $V_s$ and $V_a$; 

i) the ion density $N$, the $K_p$ index increased by coefficient 10; 

j) the solar-wind bulk velocity $V$, the $AE$ index. 

Below we discuss the dynamics of these parameters for various consequences of solar wind types. 

\subsection{CIR and IS/CIR phenomena}

Figures 1 and 2 present the average temporal profiles for sequences of events 
SW/CIR/SW and SW/IS/CIR/SW, i.e. the distinction between drawings consists in existence of 
the interplanetary shock (IS) in figure 2.
Vertical dashed lines in figures show the first (point No. 6) and last (point No. 25) points of CIR interval. 

Both figures contain the dynamics of characteristic parameters  for CIR type:
 
1. the high values of the $\beta$-parameter, thermal pressure $P_t$ and ratio of temperatures $T/T_{exp}$ 
(in comparison with undisturbed solar wind) throughout all interval,

2. the increase of bulk, sound and Alfvenic speeds throughout all interval,
 
3. the increase of density, dynamic pressure and magnitude of  magnetic field 
at the beginning of the interval with the subsequent their reduction by the end of the interval, 

4. the gradual increase of temperature and thermal pressure, 

5. the turn of the direction of stream from -2 to +2 degrees of $\phi$ angle,

6. the Alfven speed is close to sound one, 
 
7. the small average values of components of magnetic and electric fields,
 
8. the small increase in magnetospheric activity  ($D_{st}, D_{st}^*, K_p$ and $AE$ indexes)

9. the value of measured  $D_{st}$ index is higher then the value of density-corrected $D_{st}^*$. 

  
Distinctions between figures generally consist in more abrupt change of a number of parameters 
when crossing an interplanetary shock . The reason of generation of the shock can be connected 
with higher speed of the solar wind (changing from 380 to 495 and from 410 to 530 km/s) and 
higher Alfvenic Mach number (9.1 and 9.5). Main differences are following: 

1. the density, the thermal and dynamic pressures, the magnitude of magnetic field are higher for CIR with IS, 

2. the magnetospheric activity on basis of all indexes is higher for CIR with IS. 

It should be noted that the higher variability of several parameters for CIR with IS may be connected 
with lower statistics of events relative to CIR without IS.  

\subsection{Ejecta, Sheath/Ejecta and IS/Sheath/Ejecta phenomena}

Figures 3-5 present the average temporal profiles for sequences of events 
SW/Ejecta/SW, SW/Sheath/Ejecta/SW and SW/IS/Sheath/Ejecta/SW. 
Three figures have characteristic features of Ejecta: 

1. the low values of the $\beta$-parameter, thermal pressure $P_t$ and the ratio of temperatures $T/T_{exp}$ 
(in comparison with undisturbed solar wind) throughout all interval, 

2. the decrease of bulk speed throughout interval,

3. the moderate and low value of density and temperature,

4. the moderate and high magnitude of magnetic field, 


It is possible to note several features of Ejecta dynamics:

1. the decrease of temperature throughout interval,

2. the increase of density throughout interval,

3. the decrease of $\phi$ angle throughout interval,

4. the Alfven speed is higher than sound one,


Figures 3-5 have following differences for Ejecta: 

1. the bulk and Alfven speeds and the temperature are highest for IS/Sheath/Ejecta and 
lowest for Ejecta without Sheath, 

2. the magnitude of magnetic field for Ejecta without Sheath is lower and slightly increases throughout interval, 
while it for Ejecta with Sheath is higher and decreases, 

3. the Alfven and sound speeds for Ejecta without Sheath are lower and do not change throughout interval, 
while they for Ejecta with Sheath are higher and slightly decrease, 

4. the $D_{st}$ and $D_{st}^*$ indexes are close to each other for all subtypes of Ejecta, 

5. the $D_{st}$ and $D_{st}^*$ indexes for Ejecta without Sheath are $\sim$ -10 nT and do not change throughout interval, 
while they for Ejecta with Sheath are more negative and increase throughout interval 
from $\sim$ -30 up to $\sim$ -20 nT 
for Sheath/Ejecta and from $\sim$ -50 up to $\sim$ -30 nT for IS/Sheath/Ejecta case.   


Main features of Sheath in figure 4, 5 are following: 

1. the high values of the $\beta$-parameter, thermal pressure $P_t$ and the ratio of temperatures $T/T_{exp}$ 
(in comparison with undisturbed solar wind) throughout all interval,

2. the increase of bulk, sound and Alfvenic speeds throughout all interval,
 
3. the increase of density, dynamic pressure and magnitude of  magnetic field 
at the beginning of the interval with the subsequent their reduction by the end of the interval, 

4. the gradual increase of temperature and thermal pressure, 

5. the turn of the direction of stream from -2 to +2 degrees of $\phi$ angle,

6. the Alfven speed is close to sound one, 
 
7. the small average values of components of magnetic and electric fields,
 
8. the small increase in magnetospheric activity  ($D_{st}, D_{st}^*, K_p$ and $AE$ indexes)

9. the value of measured  $D_{st}$ index is higher then the value of density-corrected $D_{st}^*$. 


\subsection{MC, Sheath/MC and IS/Sheath/MC phenomena}

Figures 6-8 present the average temporal profiles for sequences of events 
SW/MC/SW, SW/Sheath/MC/SW and SW/IS/Sheath/MC/SW. 
A number of lines have not smooth form because of small statistics. 
Nevertheless, it is possible to make several conclusions. 

Dynamics of MC in figures 6-8 is close to dynamics of Ejecta in figures 3-5. 
There are following differences between MC and Ejecta. 

1. MC has higher magnitude of magnetic field and Alfven speed than Ejecta, 

2. during MC the magnetic activity is higher than during Ejecta. 

Dynamics of Sheath before MC is close to dynamics of Sheath before Ejecta. 

1. Sheath before MC has higher magnitude of magnetic field than before Ejecta, 

2. the magnitude of magnetic field is lower in Sheath before MC than in MC while 
the magnitude of magnetic field is higher in Sheath before Ejecta than in Ejecta, 

3. during Sheath before MC the magnetic activity is higher than during Sheath before Ejecta.  

\section{Discussion and Conclusions} 

First of all it should be noted that in most cases the components of the magnetic and 
electric fields are close to zero 
while the magnetospheric activity is noticeable. 
It can be explained by two facts. 
First, the components in each point can be random, and their averaging results in values close to zero. 
Second, the magnetospheric activity is not linear function of the components 
(for example, as showed in our recent empirical works 
\citep{Nikolaevaetal2013,Nikolaevaetal2015}, 
$Dst$ and $Dst^*$ indexes are well approximated by linear function of integral of electric field) 
and therefore averaging of indexes leads to nonzero result. 

In general our results on temporal profile of parameters in CIR are close 
to results previously obtained by various authors (e.g. 
\cite{BorovskyDenton2010} and references therein), 
however, unlike the previous authors, we made the analysis separately for CIR without IS and with IS. 
Our results confirm the natural dependence: 
the formation of IS before CIR is connected with higher bulk speed 
and Alfven Mach number. Generation of IS before CIR results in the increase of magnitude 
and components of IMF and therefore in the increase of the magnetospheric activity 
as illustrated by the variation of all indexes. 

Obtained results on average values and temporal profiles of parameters in ICME 
(both Ejecta and MC) are in good agreement with earlier published results 
\citep{ZurbuchenRichardson2006,RichardsonCane2012,MitsakouMoussas2014}. 
In addition to previous data we consider separately ICME with Sheath and IS, and ICME without them.  
Our results show that the formations of Sheath and then IS before ICME are connected 
with increasing bulk speed and Alfven Mach number. 
Occurrences of Sheath and then IS before ICME increase the  magnitude and components of IMF 
and therefore increase the magnetospheric activity. 

In contrast with the CIR and ICME types, the Sheath type is investigated rather poorly, 
and our analysis, apparently, is the first analysis of this sort. 
The main conclusion is that the convincing evidence is obtained 
that the behavior of parameters in Sheath and in CIR are very similar 
not only qualitatively (on the temporary profiles), but also quantitatively.

The indication in favor of a hypothesis are obtained that the speed angle $\phi$ 
in ICME changes from 2 to -2 degrees 
while it in CIR and Sheath changes from -2 to 2 degrees, i.e.  the streams in CIR/Sheath 
and ICME deviate in the opposite side. 
It can be explained by the interaction of fast ICME with slow plasma in Sheath. 

We consider that the results presented here are only the initial stage of researches: 
in the subsequent works we plan to describe in detail some interesting facts 
which are described only briefly here, and also to compare the full set of events to events 
which were geoeffective. 




%
%
%
%
%
%
%

\begin{acknowledgments}
The authors are grateful for the opportunity to use the OMNI database. The OMNI data were obtained from 
GSFC/ SPDF OMNIWeb (http://omniweb.gsfc.nasa.gov). This work was supported by the 
Russian Foundation for Basic Research, project 
13--02--00158, and by 
Program 9 of Presidium of the Russian Academy of Sciences.
\end{acknowledgments}

\end{article}


%
%

%
%
%
%
%


\begin{figure}
\noindent\includegraphics[width=14cm]{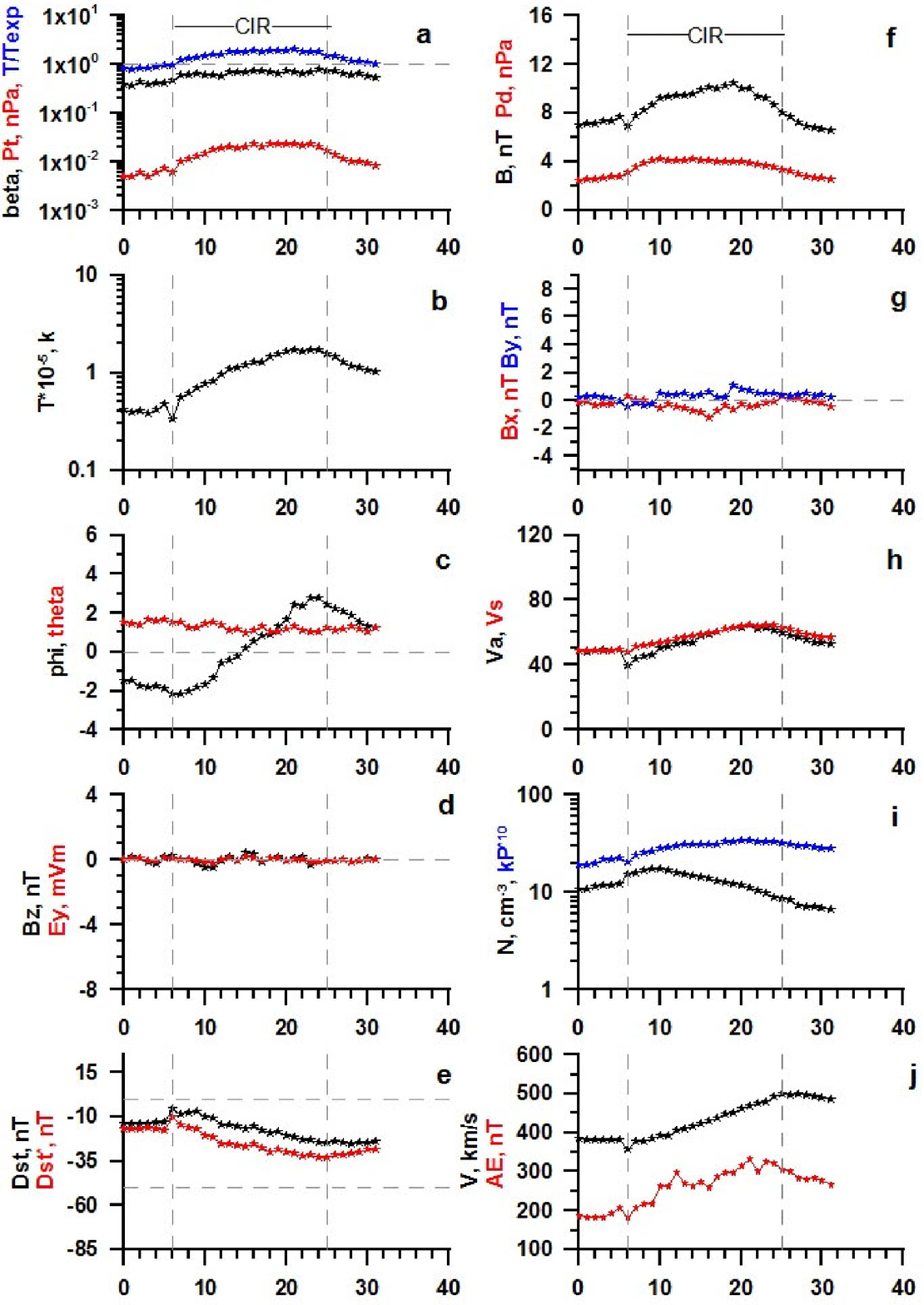}
\caption{The temporal profile of solar wind and IMF parameters for CIR 
 obtained by the double superposed epoch analysis}
\end{figure}


\begin{figure}
\noindent\includegraphics[width=14cm]{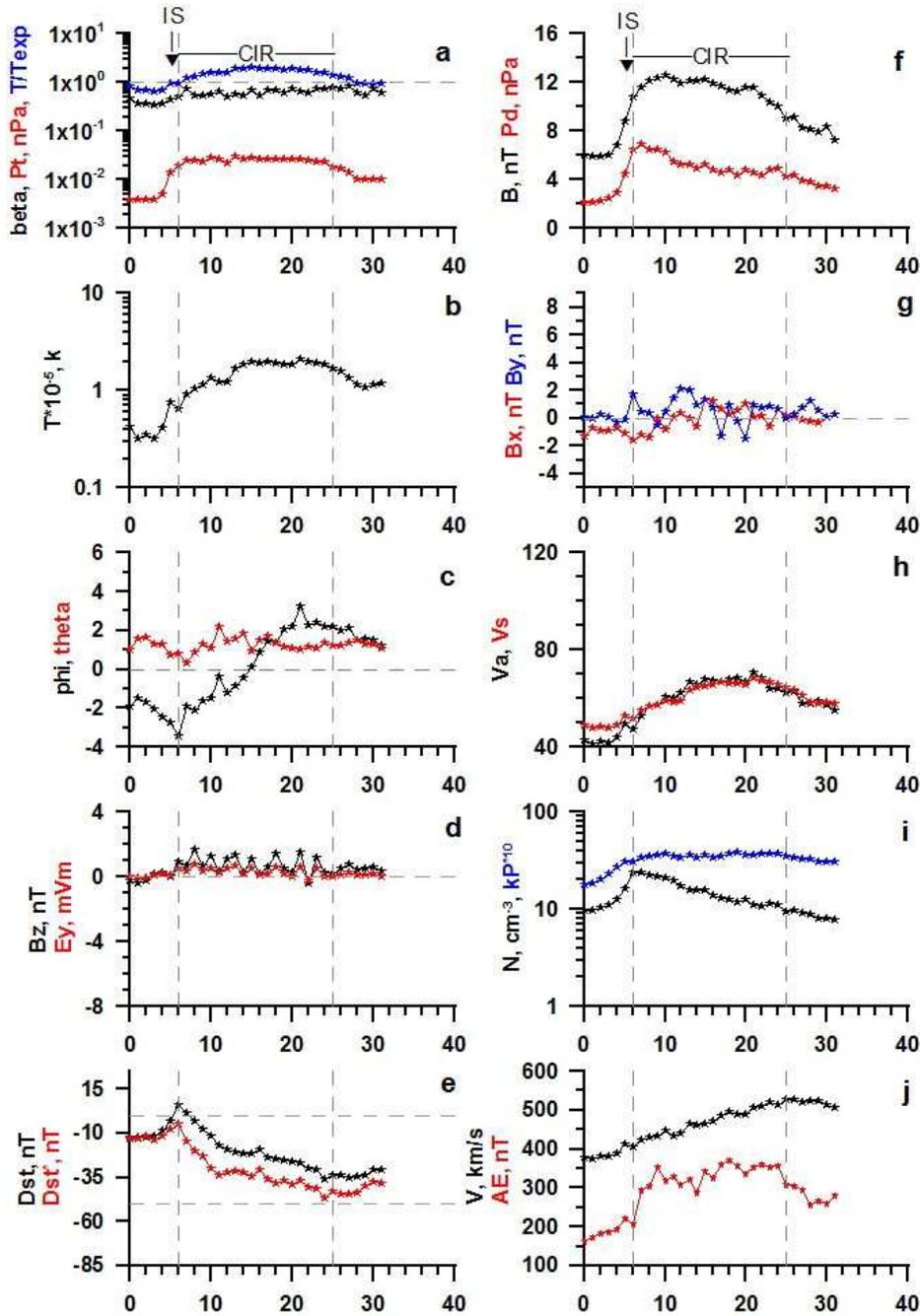}
\caption{The same as in Fig.1 for IS+CIR}
\end{figure}


\begin{figure}
\noindent\includegraphics[width=14cm]{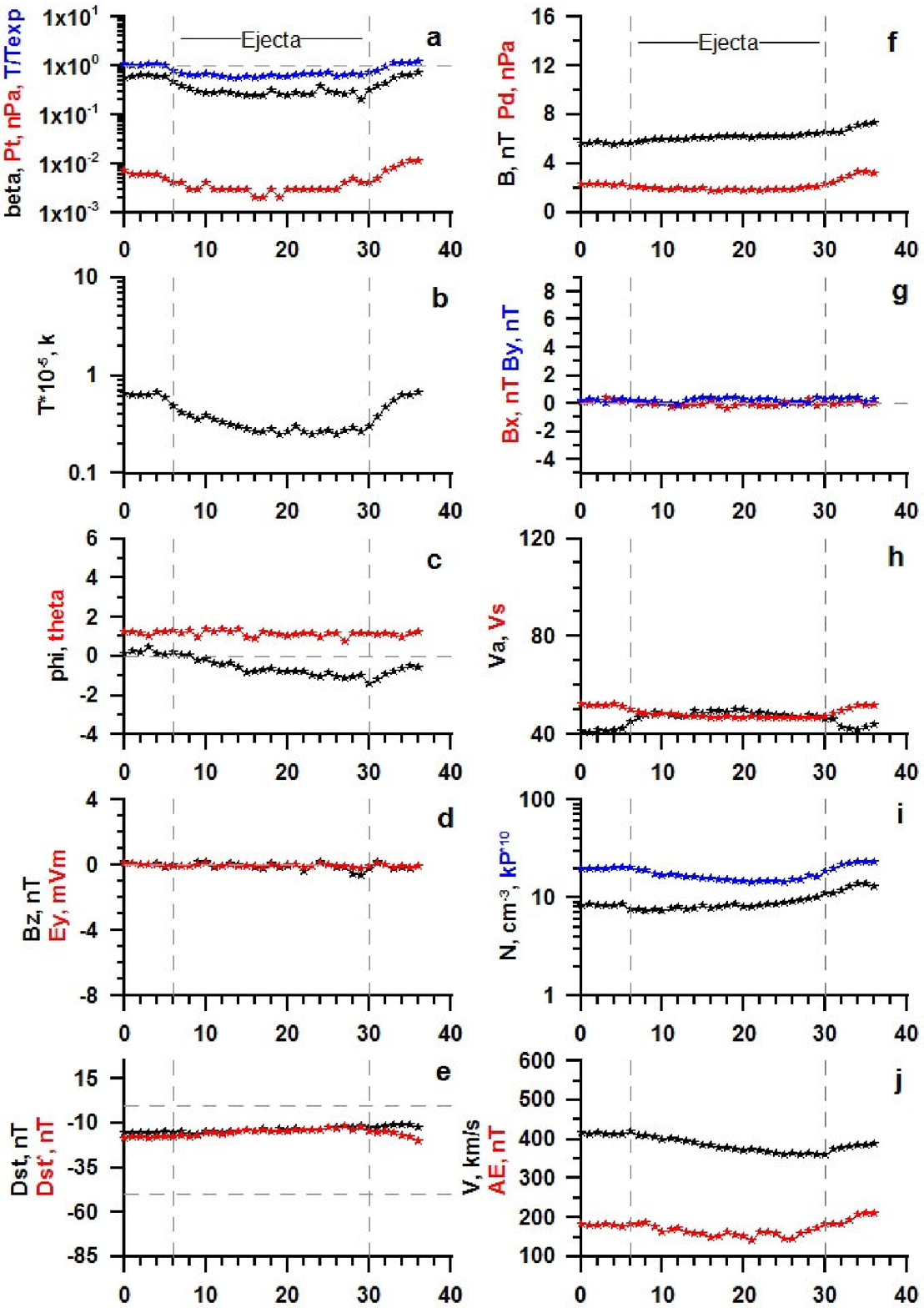}
\caption{The same as in Fig.1 for Ejecta}
\end{figure}


\begin{figure}
\noindent\includegraphics[width=14cm]{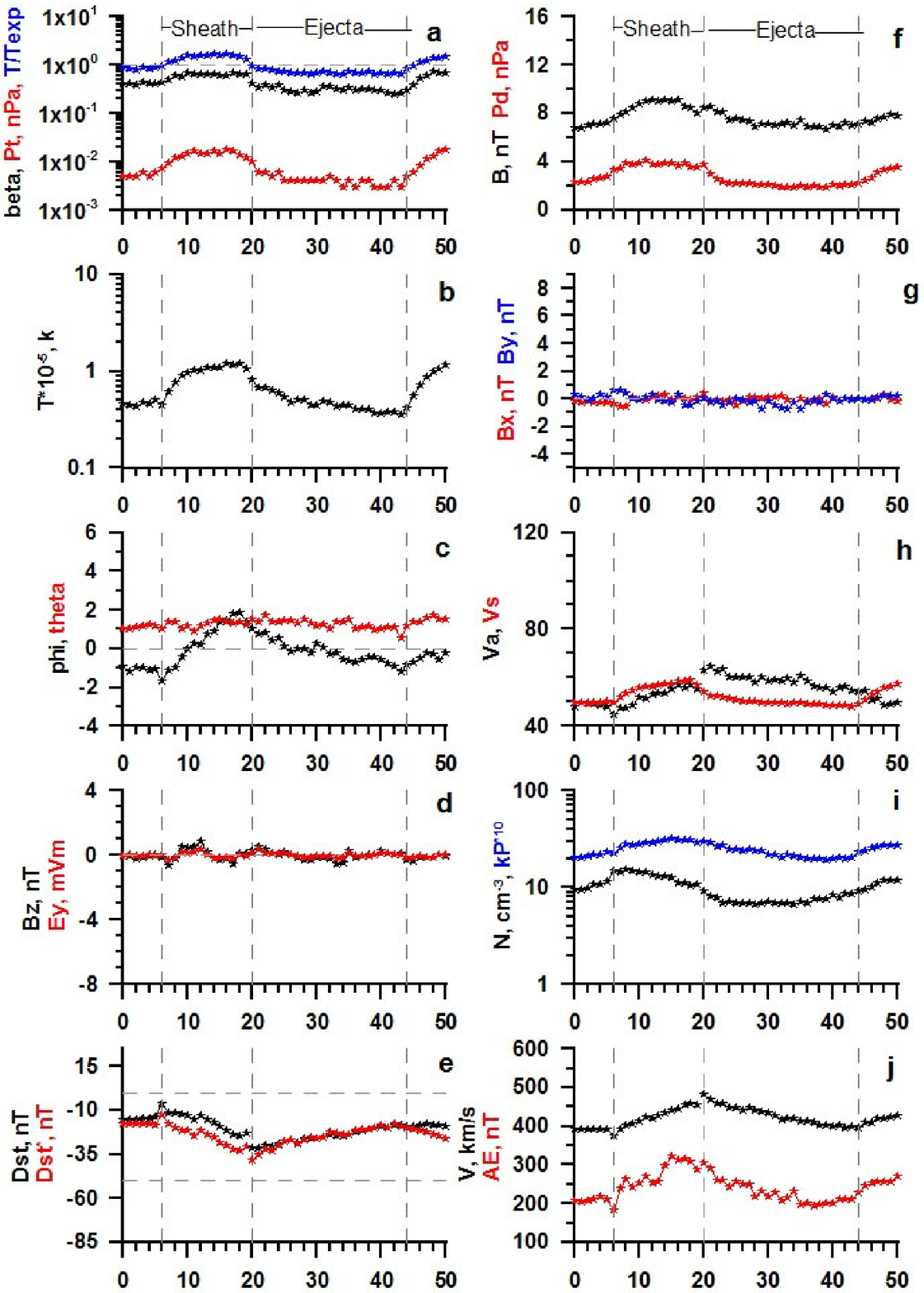}
\caption{The same as in Fig.1 for Sheath+Ejecta}
\end{figure}

\begin{figure}
\noindent\includegraphics[width=14cm]{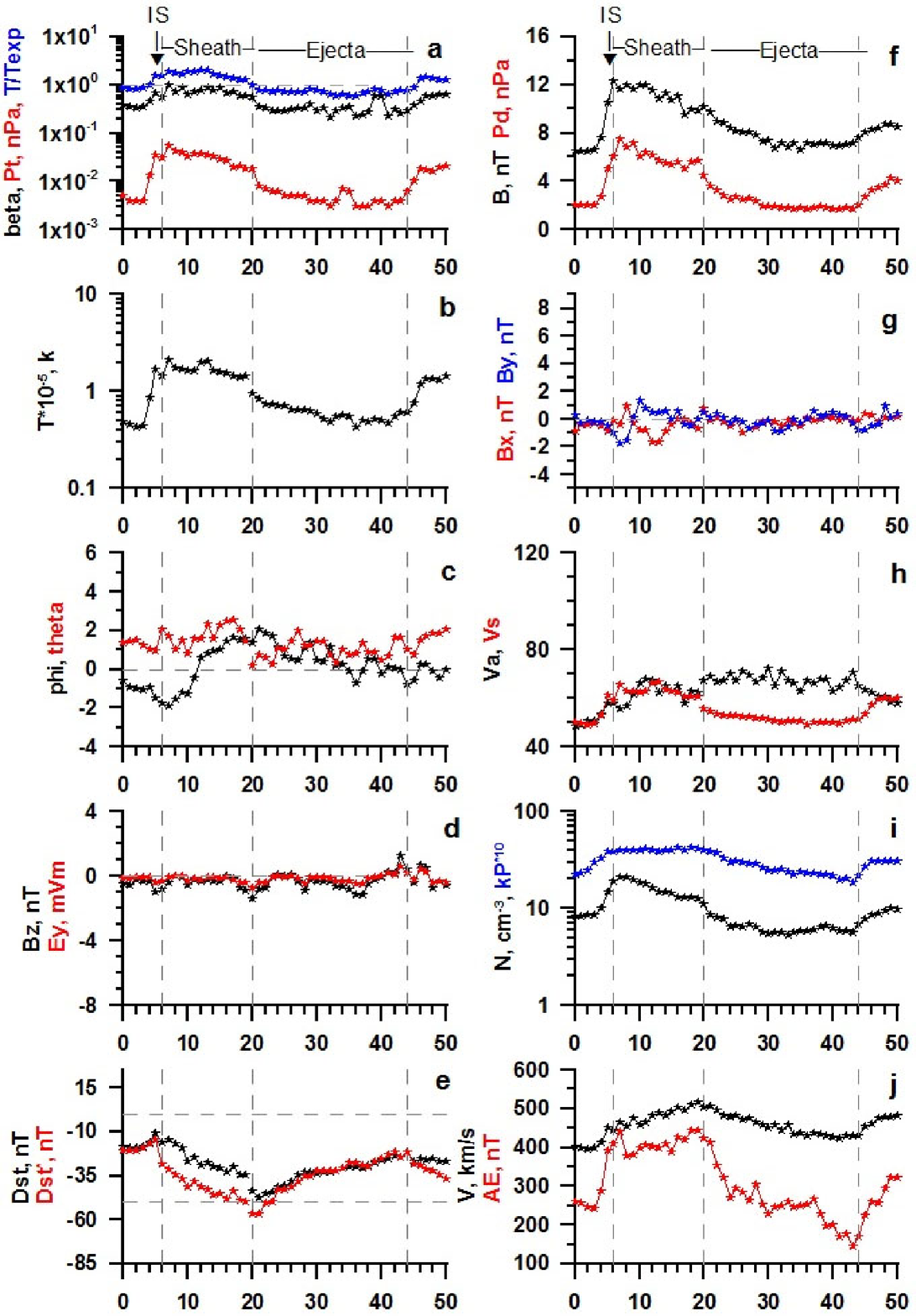}
\caption{The same as in Fig.1 for IS+Sheath+Ejecta}
\end{figure}

\begin{figure}
\noindent\includegraphics[width=14cm]{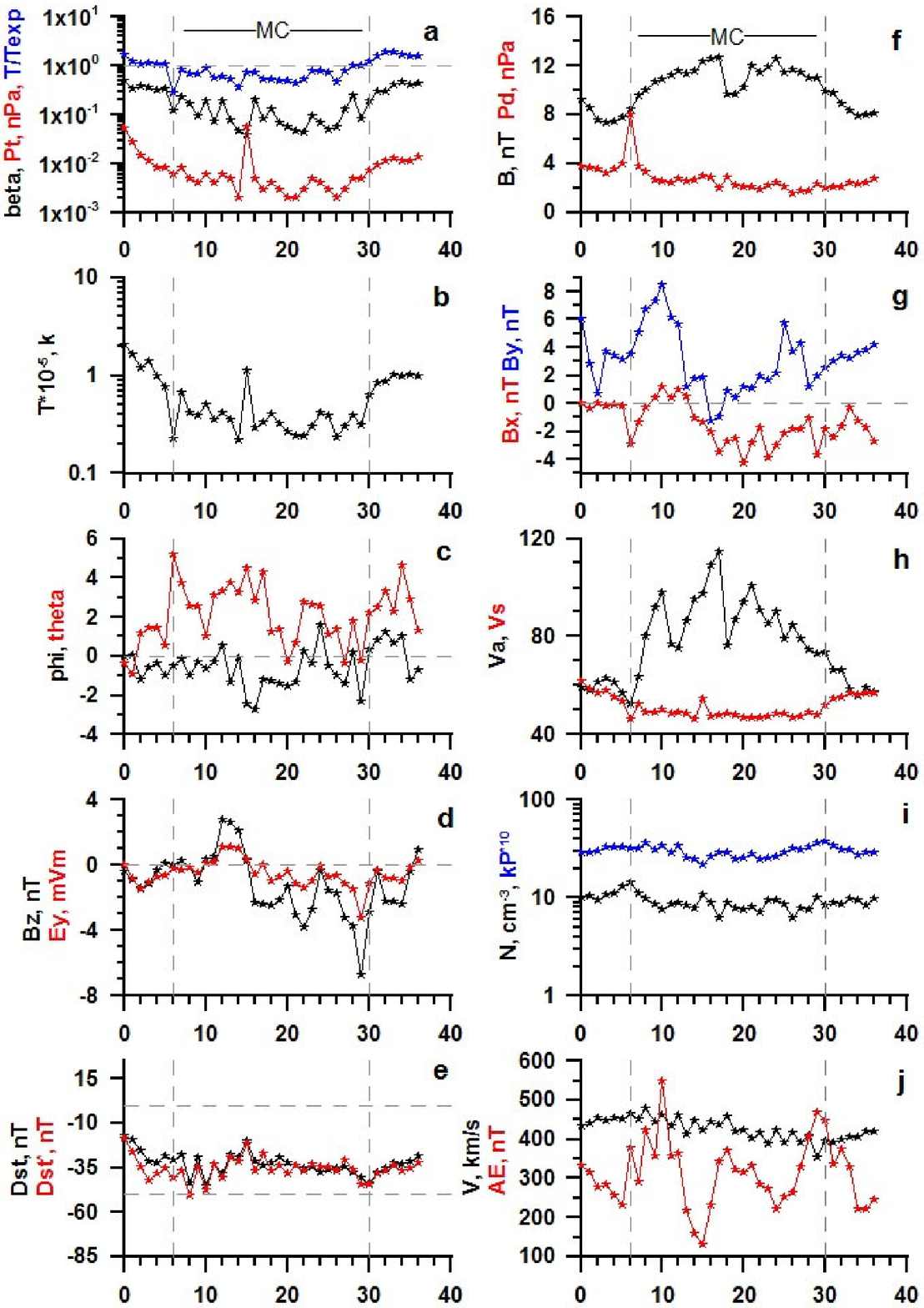}
\caption{The same as in Fig.1 for MC}
\end{figure}

\begin{figure}
\noindent\includegraphics[width=14cm]{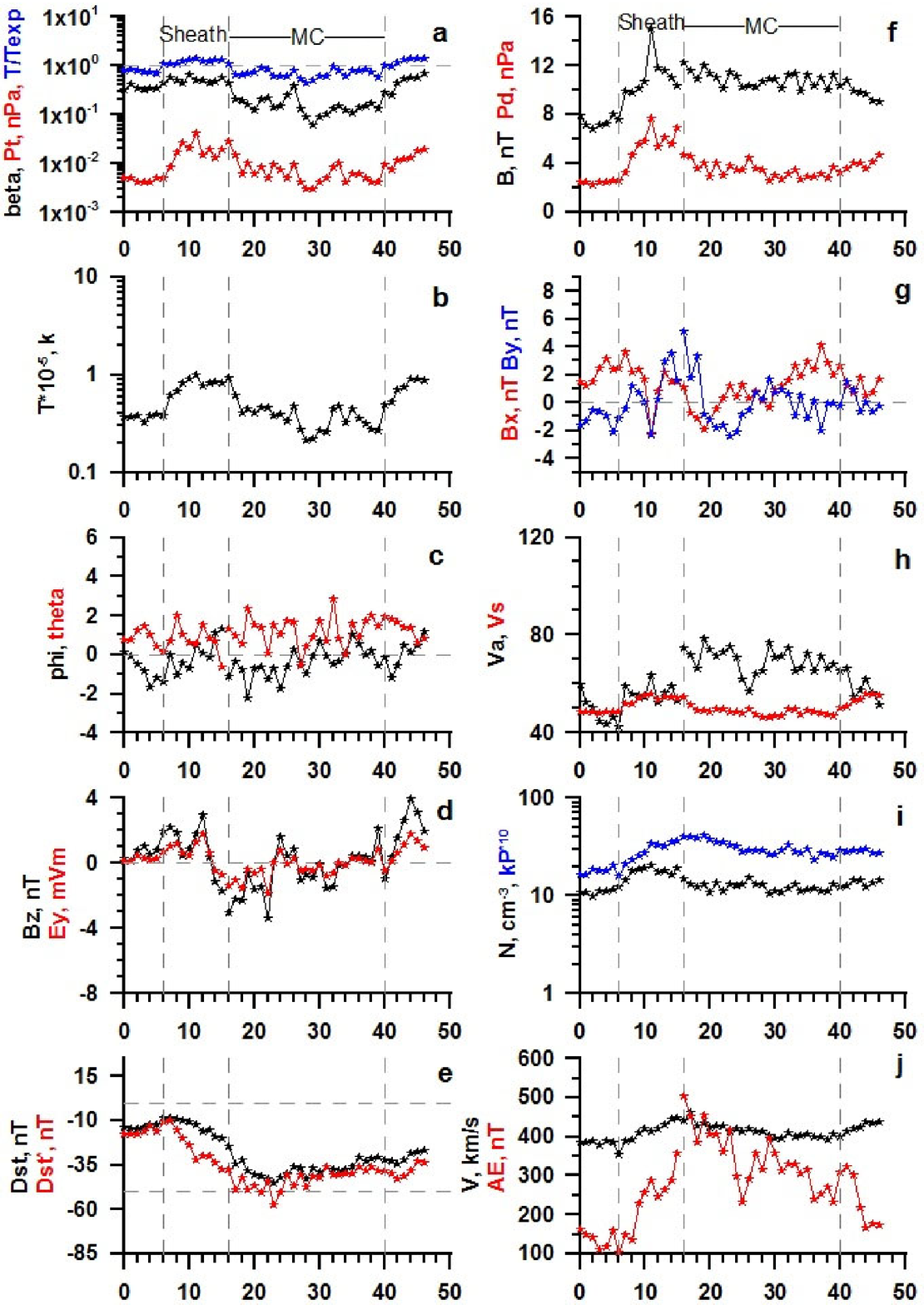}
\caption{The same as in Fig.1 for Sheath+MC}
\end{figure}

\begin{figure}
\noindent\includegraphics[width=14cm]{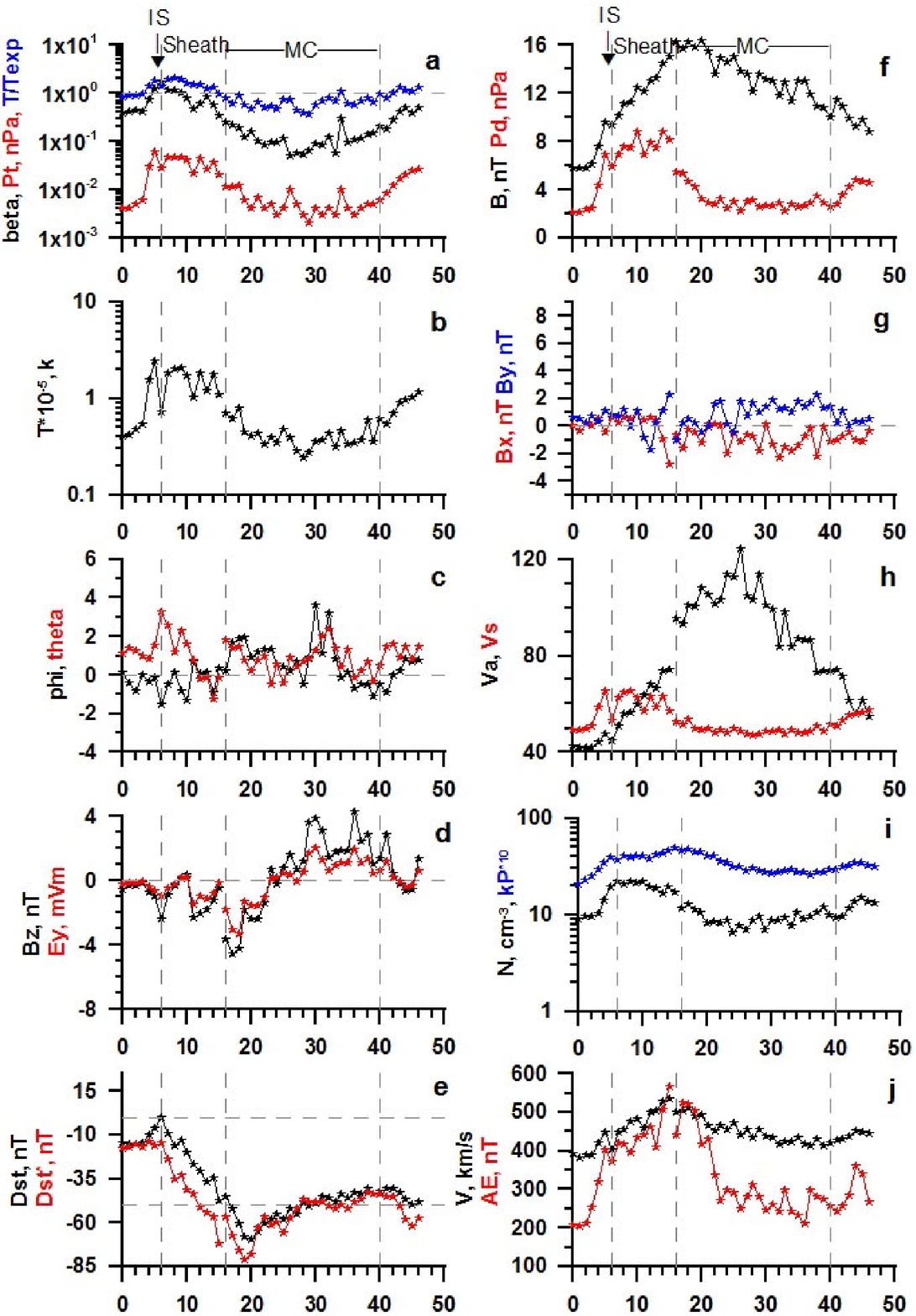}
\caption{The same as in Fig.1 for IS+Sheath+MC}
\end{figure}

%

\end{document}